\documentclass[manuscript,screen]{acmart}
\AtBeginDocument{%
  }

\setcopyright{acmlicensed}
\copyrightyear{2018}
\acmYear{2018}
\acmDOI{XXXXXXX.XXXXXXX}
\acmConference[Conference acronym 'XX]{Make sure to enter the correct
  conference title from your rights confirmation email}{June 03--05,
  2018}{Woodstock, NY}

\acmISBN{978-1-4503-XXXX-X/2018/06}

\usepackage[english]{babel}
\usepackage[utf8]{inputenc}
\usepackage[T1]{fontenc}
\usepackage{array}
\usepackage{multirow}
\usepackage{graphicx}
\usepackage{calc}
\usepackage{float}
\usepackage{multirow}
\usepackage{booktabs}
\usepackage{graphicx}
\usepackage{hyperref}
\usepackage{caption}
\usepackage{soul}
\usepackage{xcolor}
\usepackage{comment}
\begin{document}
\title{Towards Human-Centric Evaluation of Interaction-Aware Automated Vehicle Controllers: A Framework and Case Study}
\author{Federico Scarì}
\email{f.scari@tudelft.nl}
\author{Olger Siebinga}
\email{o.siebinga@tudelft.nl}
\author{Arkady Zgonnikov}
\email{a.zgonnikov@tudelft.nl}
\authornotemark[1]
\affiliation{%
  \institution{Delft University of Technology}
  \city{Delft}
  \country{The Netherlands}
}
\renewcommand{\shortauthors}{Scarì et al.}
\begin{abstract}
  As automated vehicles (AVs) increasingly integrate into mixed-traffic environments, evaluating their interaction with human-driven vehicles (HDVs) becomes critical. In most research focused on developing new AV control algorithms (controllers), the performance of these algorithms is assessed solely based on performance metrics such as collision avoidance or lane-keeping efficiency, while largely overlooking the human-centred dimensions of interaction with HDVs. This paper proposes a structured evaluation framework that addresses this gap by incorporating metrics grounded in the human-robot interaction literature. The framework spans four key domains: a) interaction effect, b) interaction perception, c) interaction effort, and d) interaction ability. These domains capture both the performance of the AV and its impact on human drivers around it. To demonstrate the utility of the framework, we apply it to a case study evaluating how a state-of-the-art AV controller interacts with human drivers in a merging scenario in a driving simulator. Measuring HDV--HDV interactions as a baseline, this study included one representative metric per domain: a) perceived safety, b) subjective ratings, specifically how participants perceived the other vehicle’s driving behaviour (e.g., aggressiveness or predictability) , c) driver workload, and d) merging success. The results showed that incorporating metrics covering all four domains in the evaluation of AV controllers can illuminate
critical differences in driver experience when interacting with AVs. This highlights the need for a more comprehensive evaluation approach. Our framework offers researchers, developers, and policymakers a systematic method for assessing AV behaviour beyond technical performance, fostering the development of AVs that are not only functionally capable but also understandable, acceptable, and safe from a human perspective.
\end{abstract}


\received{11 July 2025}
\maketitle
\section{Introduction}
Automated vehicles (AV) will be an essential part of the future of transportation~\cite{hancock2019future, Milakis}. While some vehicle automation efforts focus on scenarios where AVs operate in closed environments (e.g., warehouse vehicles), some of the most promising use cases (e.g., robotaxis, autonomous public transport) require AVs to integrate into the conventional traffic currently comprising mostly human-driven vehicles (HDVs). For this reason, much research has aimed to develop control algorithms for AVs that would allow them to interact with humans around. Traditional control approaches such as model-predictive control do not account for variability and uncertainty in human behaviour, so a new class of algorithms for AV planning and decision making has emerged recently ---  interaction-aware controllers (IACs). 

Interaction-aware controllers typically rely on a model of human driver behaviour for predicting how humans would respond to AV manoeuvres~\cite{evestedt2016interaction, sadigh2018planning,fisac2019hierarchical,schwarting2019social,siebinga2022human}. Incorporating these predictions into a cost function that prioritizes aspects like safety and comfort, the IAC then identifies the most appropriate driving policy for the AV. Increasingly many studies have been proposing new IAC architectures, with the goal of enhancing AV-HDV interactions
\cite{ban2007game,wang2015game,liu2017path,zhou2016impact,evestedt2016interaction,ward2017probabilistic,sadigh2018planning,hu2019trajectory,schwarting2019social}. However, when developing and reporting new control approaches, researchers tend to focus on AV-centred measures, evaluating, for instance, vehicle speed and acceleration profiles \cite{ban2007game, wang2015game, sadigh2018planning}, execution time \cite{wang2015game,evestedt2016interaction}, safety margins \cite{liu2017path,sadigh2018planning,schwarting2019social}, and velocity adjustments to minimize a cost function \cite{wang2015game}, with evaluations primarily conducted through simulations without involving human drivers. Critically, such AV-centric evaluation overlooks the very essence of the intended use-case for IACs -- interactions with actual human drivers. This is critical because the integration of AVs into the existing road system relies to a large extent on the acceptance by the drivers of HDVs around them~\cite{vinkhuyzen2016developing, nastjuk2020drives, domeyer_vehicle_2020}. 
Acceptance of AV by a human in this context encompasses a variety of interrelated concepts, including perceived ease of use, attitude, adherence to social norms, trust, usefulness, risk, compatibility \cite{jing2020determinants} as well as the legibility of AV behaviour \cite{lichtenthaler2012increasing}. All of these aspects often remain overlooked in the current literature reporting increasingly many IACs for different interactive scenarios but leaving out actual interactions from the evaluation.

Adjacent fields such as human-robot interaction and traffic psychology have begun to develop an understanding of how humans interact with AVs, but the development of IACs has so far remained largely disconnected from these fields. Furthermore, the current literature on human-AV interactions is mostly limited to a) studies of human responses to ``dummy'', pre-programmed AVs in driving simulators (e.g.,~\cite{miller_implicit_2022, zgonnikov_nudging_2024, ma_driver-automated_2024}) which exclude AVs with interactive capabilities, and b) case-based observations of human interactions with real-world AVs (e.g., ~\cite{brown_halting_2023}) which lack fine-grained information on the dynamics of the interaction and human perception of it. In summary, there is a substantial gap between the development of interaction-aware AV controllers and studies of human-AV interactions.

To address this gap, here we propose a conceptual framework for evaluating interactive capabilities of AV controllers in interactions with real humans. This framework is grounded in previous literature on human-robot interaction, and aims to provide researchers who develop AV controllers with guidelines on evaluating their algorithms in a way that captures the intended use case. We illustrate our framework with a case study focusing on evaluating interactions of human participants with one of the most popular IACs~\cite{sadigh2018planning} in a highway merging scenario, using HDV--HDV interactions as a baseline.

\subsection{Related work}
\subsubsection{Evaluating interactive capabilities of AV controllers}
The literature reporting IAC algorithms is vast, spanning a variety of control approaches and target scenarios~\cite{schwarting_planning_2018,di_survey_2021,siebinga2022human,wang_social_2022}. The main focus of a typical paper in the field is, understandably, on developing a new control algorithm. While evaluation is almost always part of such a paper, it is typically limited, especially when it comes to the more complex aspects of controller performance such as interaction with humans.

One common approach that circumvents the complexities associated with evaluating IAC interactions with humans is evaluating the IAC in scenarios with \textit{simulated} human agents. Such simulated agents can be either non-reactive (e.g., based on real-world trajectories) or reactive (based on a human driver model). In the former case, the use of human-generated trajectories as an input to the controller (e.g., \cite{liu2022interaction,li_interaction-aware_2024}) is supposed to ensure the realism of the scenario. However, the fact that surrounding agents do not react to AV's actions precludes meaningful assessment of the controller's interactive capabilities. In the latter case, this issue is alleviated by simulated surrounding agents in a closed loop via a human model (e.g.,~\cite{evestedt2016interaction,fisac2019hierarchical,schwarting2019social,liu2022interaction}). However, such model-based evaluations can only provide proxy measures of the actual AV-HDV interactions because they still do not involve actual HDVs. Further limiting this approach, simple models commonly used in the field (such as reward optimization or intelligent driver model) are known to misrepresent human driver behaviour~\cite{siebinga2022human, durrani_new_2024}. Finally, neither the open-loop trajectory-based nor the closed-loop model-based approach can provide information on humans' subjective perception of interactions.

For these reasons, an increasing number of studies have begun to evaluate AV controllers in scenarios involving actual human drivers. Yet, such evaluations tend to remain rudimentary. In most cases, the aim is to demonstrate the feasibility of online operation of the controller rather than to systematically examine the resulting AV-HDV interactions.

For instance, Yu et al.~\cite{yu2018human} and Coskun et al.~\cite{coskun2019receding} evaluated lane-changing controllers in human-in-the-loop driving simulators, using first- and third-person perspectives, respectively. However, both studies only report results from single interaction trials and offer no in-depth analysis of the interaction process. Similarly, Zhang et al.~\cite{zhang2019game} involved 20 participants in a third-person-view driving simulator study of their controller in lane changes, but only reported interaction dynamics for a few illustrative examples, focusing the main analysis on estimating driver aggressiveness.

The study by Sadigh et al.~\cite{sadigh2018planning}, which introduced one of the most influential IACs to date, included eight participants interacting with the controller in lane-changing and intersection scenarios. In a top-down-view driving simulator, the participants interacted with an AV in either attentive or distracted driving conditions. The evaluation focused mostly on differences between how IAC interacted with attentive vs. distracted drivers based on qualitative visual inspection of averaged trajectories, and only reported systematic analysis for coarse behavioural outcomes (i.e., whether the human yielded to the AV).

More recent efforts have begun to adopt more elaborate evaluation methods. For example, Hang et al.~\cite{hang_brain-inspired_2023} used a first-person, multi-agent driving simulator in a lane change task with asymmetric roles: an IAC or one of the two human participants initiated the lane change, while three other human participants occupied the adjacent lane and responded to the manoeuvrer. The evaluation included metrics such as trajectory similarity, lane change velocity, and a computed aggressiveness score based on vehicle dynamics. However, the emphasis remained on comparing IAC behaviour to that of the two humans performing the same task, rather than analysing the interaction between AVs and human drivers.

One of the most comprehensive IAC evaluation studies to date is reported by Li et al.~\cite{li_safe_2023}, who examined interactions of their IAC at intersections with 24 participants. Importantly, their study systematically analysed participants' subjective perceptions of interaction quality, such as safety and comfort, and complemented these with EEG measures of driver state. Yet even in this case, each participant encountered the IAC in only three trials (one per speed limit condition), and the analysis of interaction process was limited to a few representative examples.

Overall, while human-in-the-loop evaluations of AV controllers are becoming more common, current practices do not allow researchers to capture the potentially rich and variable nature of AV-HDV interactions. Analyses typically focus on high-level outcome measures such as interaction priority and rarely measure the details of how AV-HDV interactions unfold over time as well as participants' subjective experiences in these interactions.

\subsubsection{Evaluating human driver responses to simulated and real-life AVs}
In parallel with the development of IACs, detailed understanding of human behaviour around AVs has been the subject of increasingly much research. Such research mostly centres around \textit{vulnerable road users} (VRUs), with perhaps the most studied scenario being pedestrian crossing, where the effects of implicit and explicit communication by AVs have been extensively investigated (e.g.,~\cite{dey_pedestrian_2017, habibovic_communicating_2018, razmi_rad_pedestrians_2020, rasouli_autonomous_2020, de_winter_external_2022, zach_noonan_kinematic_2023, harkin_how_2024, lee_hello_2025}). Such studies, however, tend to use pre-recorded, non-interactive AV trajectories, with only few exceptions; for instance, Camara et al. investigated pedestrian interactions with a simulated autonomous vehicle controlled by a game-theoretic IAC in virtual reality~\cite{camara_evaluating_2021}.

Studies that aim to understand how human \textit{drivers} interact with AVs are more scarce, but span a range of scenarios including car following~\cite{rahmati_influence_2019, zhao_field_2020}, deciding who goes first at an intersection~\cite{imbsweiler_cooperation_2018, zgonnikov_nudging_2024, ma_driver-automated_2024}, overtaking on rural roads~\cite{mohammad_drivers_2024}, and negotiating a narrow passage~\cite{rettenmaier_matter_2021, miller_implicit_2022,yang_investigating_2025}. Compared to the literature on VRU behaviour which extensively covers both simulated and real-world scenarios, the work on AV-HDV interactions typically focuses on driving simulator studies, presumably due to the risks and complexity associated with systematically studying human drivers' interactions with real AVs. At the same time, even in simulated AV-HDV interactions, researchers have so far predominantly employed ``AVs'' with simplistic, pre-programmed behaviours instead of interactive AVs that react to HDV's behaviour.

In a related stream of research, studies involving the ``Wizard-of-Oz'' deception paradigm illuminated human responses to ``automated'' vehicles (e.g.,~\cite{habibovic_evaluating_2016,soni2022behavioral}) which were in fact driven by a concealed human driver. Similarly, ``reverse Wizard-of-Oz'' paradigms deceive participants into believing that a pre-programmed vehicle in a driving simulator is driven by a human~\cite{mohammad_drivers_2024,yang_investigating_2025}. Even though insights from these studies are valuable for understanding participants' behavioural adaptations in the presence of (apparent) AVs, they do not expose participants to actual AV behaviours and thus provide only limited insight into the interactions between HDVs and actual AVs. Such interactions have recently been explored from the ethnographic perspective by Brown et al., who studied the reactions of HDVs to a real-life AV based on publicly available video recordings~\cite{brown_halting_2023}. While bringing much-needed ecological validity and nuance to the body of knowledge on AV-HDV interactions, due to the focus on individual interaction instances such ethnographic studies do not allow for systematic evaluation of interactions.

\subsubsection{Limitations of the related work}
The studies reviewed above highlight a critical gap: despite being designed to handle interactions, current IACs are rarely assessed from the perspective of the human drivers they are meant to interact with. In the cases when such assessment is performed, it usually remains superficial in terms of the range of studied aspects of human behaviour. At the same time, studies that do aspire to provide in-depth understanding of human behaviour in interactions with AVs paradoxically tend to focus on interactions with ``dummy'', non-interactive AVs which at best signal their intention but do not react to human responses. The main contribution of this paper -- a systematic framework for evaluating interactive capabilities of automated vehicle controllers --aims to bridge this gap.

\section{The Framework for Evaluating Interactive Capabilities of Automated Vehicle Controllers}
\begin{figure*}[h]
    \centering
    \captionsetup{justification=raggedright}
    \includegraphics[width=1\textwidth]{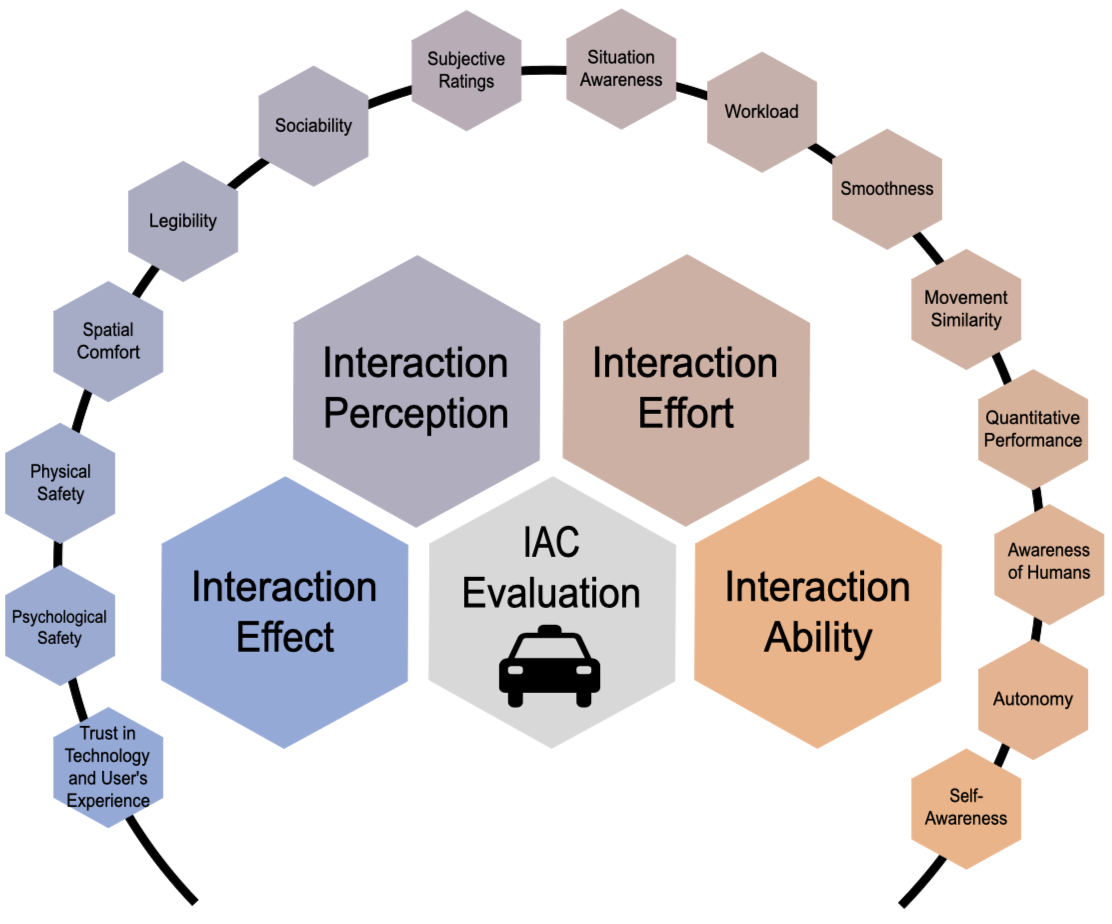}
    \caption{Proposed Evaluation Framework for Interaction-Aware Controller (IAC). The dimensions encompass metrics and methods to evaluate interactions between humans and robots, which can be applied to AVs. These dimensions form a spectrum ranging from human-centric metrics (blue), which focus on how AV behaviour is perceived and experienced by human drivers, to robot-centric metrics (orange).)}
    \label{framework}
\end{figure*}

Our framework for the Interaction-Aware Controller (IAC) evaluation consists of four domains: interaction perception, effort,  effect, and ability (Figure~\ref{framework}). Each domain is based on the literature from the human-robot interaction (HRI) research fields. The domains encompass metrics and methods (dimensions) to evaluate interactions between humans and robots, which can be applied to AVs. These dimensions form a spectrum ranging from human-centric metrics to robot-centric metrics, assessing the AV's autonomous capabilities, such as self-awareness, human awareness, and movement similarity. Since the manifestation of each dimension may differ based on the use case or research goal, the order of dimensions on this spectrum can change. However, the spectrum is meant as a guiding structure to reach the common shared goal: evaluating the interaction between AVs and HDVs.

The dimensions have been grouped into four domains to further structure and simplify the framework. Some metrics naturally influence multiple domains because of the domains' interconnected nature. However, this doesn't diminish the benefit of using the domains as they allow for a structured evaluation. By evaluating each domain separately, using some --but not necessarily all-- of the underlying dimensions, we can ensure a holistic evaluation of the interaction, balancing between human experience and technological performance. Thereby, the frame incorporates the importance of evaluating both human and technological factors to ensure efficiency, acceptance, \underline{and} safety.

\subsection{DOMAIN 1: Interaction Effect}
The Interaction Effect domain focuses on how the interaction between AVs and HDVs influences the human driver’s experience. This domain emphasizes the emotional and cognitive impact of the interaction, assessing whether the AV's behaviour fosters trust, comfort, and a sense of safety in the human driver. Additionally, this domain considers the driver’s psychological priming before interacting with the AV—how their expectations and initial mindset influence (has an effect on) the interaction itself.
It represents the most human-centric and subjective perspective in the framework and forms the foundation for understanding the driver’s experience.

\subsubsection{Trust in Technology and User Experience}
Trust in technology and the user’s experience with automation are critical factors influencing the acceptance and effective use of autonomous systems. This is particularly important in the context of autonomous vehicles, where user perceptions and interactions can vary significantly. Unlike a one-size-fits-all approach, researchers should strive for flexible, context-sensitive automation that acknowledges individual differences in how users interact with technology.
A closely related concept is the user’s attitude toward using technology, which can significantly impact their trust levels. The Unified Theory of Acceptance and Use of Technology (UTAT) \cite{venkatesh2003user} framework provides a structured way to evaluate this attitude, defining it as the sum of all positive or negative feelings about completing tasks supported by automation \cite{weiss2009usus}. By incorporating UTAT into evaluations, researchers can distinguish between participants with high trust and those with lower trust in automation, offering deeper insights into individual differences.
Moreover, combining trust assessments with other metrics (e.g., such as workload evaluations through questionnaires like NASA-TLX), can provide a nuanced understanding of how trust interacts with cognitive demands during AV-HDV interactions. For instance, a lower level of trust in technology might correlate with a higher reported workload, revealing key areas for improvement in AV design and user experience.

\subsubsection{Psychological Safety}
Psychological safety refers to perceived safety during interactions, often linked to the user’s stress levels. It is a critical aspect of comfort in human-robot interaction, as it significantly influences both the user’s attitude toward technology and their trust in automation \cite{rubagotti2022perceived}. In the context of AV evaluation, psychological safety is crucial because it directly influences driver trust, acceptance, and willingness to share the road with AVs, impacting overall traffic flow and integration into human-driven environments.
Psychological safety is commonly assessed through questionnaires that focus on three main aspects: (a) “whether the interaction was free from obstruction”, (b) “whether the person could maintain their preferred velocity in the presence of the robot”, and their (c) “overall impression of the encounter” \cite{gao2022evaluation}. Examples of such questionnaires include: the Godspeed questionnaire \cite{bartneck2009measurement}, the Robotic Social Attributes Scale (RoSAS) \cite{carpinella2017robotic} (based on the Godspeed questionnaire), and the BEHAVE-II instrument \cite{joosse2013behave}.
While questionnaires are the most widely used method, other approaches have been proposed to capture a more comprehensive picture of psychological safety:
\begin{enumerate}
    \item \textit{Physiological Signals}: Tracking real-time data such as heart rate, heart rate variability, and eye gaze. These are considered reliable biomarkers of stress and fear, with heart rate being particularly significant as it reflects activation of the autonomic nervous system \cite{rubagotti2022perceived}.
    \item \textit{Behavioral Assessment}: Observing physical responses, though this may be less relevant for the more subjective aspects of perceived safety.
    \item \textit{Direct Input Devices}: Tools like joysticks allow participants to record their perceived safety in real-time during an experiment. While valuable for linking feedback to specific moments, these devices can be distracting or invasive for the user.
\end{enumerate}
The literature suggests that a combination of methods yields the most robust results \cite{rubagotti2022perceived}. For example one could use questionnaires post-interaction to capture the overall experience and complement these with real-time physiological measures like heart rate or eye-gaze tracking for insights into immediate reactions during specific moments of the interaction.
This multimodal approach ensures a thorough understanding of psychological safety, allowing researchers to evaluate how an autonomous system’s behaviour influences the user’s perceived safety and overall experience.

\subsubsection{Physical Safety}
Physical safety concerns all intentional or unintentional contact between the user and the robot. Usually, in simulations, this parameter is quantified by the number of collisions. To avoid any possible harm due to collisions, many experiments approximate collision by the invasion of a defined safety zone, which is typically derived from the proxemic theory. Physical safety is a fundamental requirement for AV integration, as ensuring reliable collision avoidance and risk mitigation is essential for gaining public trust and regulatory approval.

\subsubsection{Spatial Comfort}
Spatial comfort refers to the physical and interpersonal space maintained during interactions, which significantly influences user acceptance and comfort. While psychological comfort is challenging to quantify, researchers often focus on physical comfort and interpersonal space as key contributors to discomfort.

Proxemic Theory is a widely used theory that defines personal space as layers of concentric circles, each structured by their social functions. As reported by Gao et al. \cite{gao2022evaluation}, proxemic theory explains how individuals maintain space for different social purposes in interactions, providing a framework for assessing spatial comfort.
Another theoretical foundation of comfort is the Social Force Model (SFM). SFM is originally developed to simulate human navigation in social contexts, the SFM quantifies comfort by modelling repulsive forces from nearby agents and obstacles. It can also highlight violations of social comfort in scenarios such as emergencies, where panic behaviours occur \cite{gao2022evaluation}.

Spatial comfort can be assessed using:
\begin{enumerate}
    \item \textit{Proxemic Models}: Evaluate spatial violations based on distance thresholds defined by proxemic theory.
    \item \textit{Social Force Model}: Quantify the influence of repulsive forces on comfort during navigation.
    \item \textit{Social Individual Index (SII) and Social Sensitivity Index (SSI)}: Metrics that evaluate spatial comfort and sensitivity to interpersonal space during interactions.
\end{enumerate}
Spatial comfort is fundamental for AV acceptance. Safety distances are often embedded into AV controllers based on regulations and conventions, ensuring basic physical comfort. However, it is important to assess whether human drivers perceive these distances differently when interacting with an AV, as compared to another human-driven vehicle. This evaluation can reveal whether AV controllers need to adapt to driver preferences to enhance comfort and trust. By evaluating spatial comfort, researchers can ensure that AVs navigate in a manner that respects human drivers' interpersonal space, fostering a more intuitive and positive interaction experience.\hfill \break
\hfill \break
The Interaction Effect domain evaluates how well the AV facilitates a positive human experience during interactions, from the driver’s initial priming to their perceived psychological and physical safety, and trust during the interaction. Metrics in this domain ensure that the human-centric perspective is thoroughly addressed, which is crucial for the successful integration of AVs into mixed-traffic environments. By focusing on both pre-interaction (priming) and interaction experiences, this domain provides a holistic understanding of the driver’s journey.

\subsection{DOMAIN 2: Interaction Perception}
The Interaction Perception domain focuses on how human drivers perceive and interpret the behaviour of AVs. This domain emphasizes the subjective experience of the interaction, assessing whether the AV’s actions are understood, predictable, and socially appropriate. A clear and intuitive perception of AV behaviour is critical for ensuring smooth and safe interactions with HDVs.

\subsubsection{Legibility}
Legibility, sometimes referred to as readability, is a critical parameter for evaluating human-like trajectories. According to Lichtenhäler et al., \cite{lichtenthaler2012increasing}, "Robot behaviour is legible if a human can infer the next actions, goals, and intentions of the robot with high accuracy and confidence." This metric focuses on how clearly and predictably an AV communicates its intentions to human drivers, ensuring smoother and safer interactions.
Legibility is typically evaluated using subjective feedback from participants, both during and after the experiment. During the Experiment participants are asked to predict the AV's next motion and whether it would affect their trajectory. In addition, they are asked to rate their confidence in their predictions.
After the Experiment participants rate whether the AV’s actual behaviour matched their expectations and, if not, how surprising it was.
Lichtenhäler et al. recommend using a five-point Likert scale to assess these ratings, providing quantitative insights into participants' comprehension and confidence.
Legibility offers valuable insights into the human driver’s understanding of the AV's movements and intentions. This metric is particularly important for ensuring that AV behaviour is intuitive, reducing confusion and potential safety risks during interactions.
While the assessment of legibility during the experiment provides immediate insights, it can be invasive and disrupt the driver’s natural behaviour, especially in real-world scenarios. To mitigate this, post-experiment assessments can be used to evaluate legibility without interfering with the driving experience. Such assessments remain effective for capturing participants’ overall perceptions and ratings.

\subsubsection{Sociability}
Sociability, as defined in the literature, refers to a robot's ability to conform to complex social conventions. In the context of AVs, sociability encompasses adherence to social and cultural etiquettes, such as observing the right-of-way. While an AV may navigate naturally and appropriately from a technical perspective, violating expected social norms can still lead to discomfort and negative perceptions among human drivers. As noted by Kruse et al. \cite{kruse2013human}, "discomfort is caused by the violation of a social rule, rather than a violation of a comfort distance."
Quantifying sociability is one of the most difficult challenges in human-aware navigation \cite{gao2022evaluation} due to its subjective and culture-specific nature. For example, in some countries, it is customary to let vehicles merge onto highways (e.g., Germany), while in others, such practices may not be as prevalent.
This cultural variability complicates the standardization of metrics for assessing sociability in AV-HDV interactions.

While there are no standardized questionnaires specifically tailored to AV-HDV interactions, existing tools in HRI research can provide valuable insights like the Perceived Social Intelligence (PSI) Scale \cite{gao2022evaluation}, which evaluates sociability across 20 aspects or the Social Competence (SOC) Scale \cite{gao2022evaluation}, which focuses on four key items: 1) social competence, 2) social awareness, 3) social insensitivity (reversed), and 4) strong social skills.
Sample questions in these questionnaires include: "Is the robot’s behaviour socially appropriate?", "Is the robot’s behaviour friendly?", and "Does the robot understand the social context and interaction?"

Sociability is a critical factor for ensuring human acceptance of AVs in mixed-traffic environments. By evaluating whether an AV behaves in a socially appropriate manner, researchers can identify gaps in its interaction design that might hinder its integration into diverse traffic systems. The PSI and SOC scales offer structured approaches to evaluate sociability, though they must be applied with cultural considerations in mind.

For AV-HDV interactions, PSI is particularly useful, despite its limitations, as it provides a framework to assess how human drivers perceive the AV’s adherence to social norms. Incorporating sociability metrics into AV evaluations helps ensure that these systems align with both technical and social expectations, fostering safer and more intuitive interactions.

\subsubsection{Subjective Ratings}
Subjective Ratings (SR) \cite{steinfeld2006common} refer to the quality and quantity of effort a user applies to successfully interact with a robot or AV. These ratings provide a qualitative overview of the interaction, offering valuable insights into the user’s experience and perceived challenges.

Subjective ratings are typically gathered through post-experiment questionnaires, where participants evaluate their interaction experience. These questionnaires should focus on the effort required to interact with the AV effectively and the ease or difficulty of understanding and responding to the AV’s behaviour.
While SR provides a rough initial overview of the interaction, it’s important to acknowledge that subjective answers can vary significantly between participants, reflecting individual perceptions and biases.
Subjective ratings are a valuable starting point for evaluating AV-HDV interactions, offering a quick and qualitative measure of user experience. While subjective by nature, these ratings help highlight areas where drivers perceive higher effort or complexity, guiding further investigation into interaction dynamics.\hfill \break

The Interaction Perception domain highlights the human-centric viewpoint of AV-HDV interactions, ensuring that the AV’s behaviour aligns with human expectations. Metrics in this domain focus on the clarity, predictability, and social appropriateness of AV actions, which directly influence trust, acceptance, and safe interactions.

Evaluating perception is crucial for designing AVs that are not only technically efficient but also intuitive to interact with in real-world, mixed-traffic environments. By addressing how human drivers interpret AV behaviour, this domain bridges the gap between autonomous decision-making and human understanding, creating a more seamless interaction experience.

\subsection{DOMAIN 3: Interaction Effort}
The Interaction Effort domain focuses on the mental and physical workload required by human drivers to interact effectively with AVs. This domain evaluates how much cognitive and physical energy drivers need to expend during interactions and aims to identify potential stressors or inefficiencies in the interaction process.

\subsubsection{Situation Awareness}
Situation Awareness (SA) \cite{steinfeld2006common, scholtz2003theory} refers to the operator’s ability to:
\begin{enumerate}
    \item Perceive the elements in the environment at a specific time and space.
    \item Comprehend the meaning of the perceived elements.
    \item Anticipate the future status of these elements.
\end{enumerate}
This three-step process, as defined by Endsley  \cite{endsley1988situation}, is critical for effective decision-making and interaction, particularly in dynamic environments like driving. For AV-HDV interactions, SA is a key metric for understanding how quickly and accurately human drivers interpret AV behaviour.
One of the most used ways to measure SA is the Situation Awareness Global Assessment Technique (SAGAT) \cite{endsley1995measurement}. Applied to experiments, SAGAT uses the freezing technique to assess one’s SA. The SAGAT consists of freezing the experiment at a random moment and assessing the participant’s knowledge with questions about the experiment. Another method to measure SA is the Situational Awareness Rating Technique (SART) \cite{endsley1998comparative}, in which participants are asked to self-report their SA through questionnaires. Other online methods to measure SA are eye-tracking \cite{rubagotti2022perceived},
performance-based methods, and think-aloud methods.
SA is a vital metric for evaluating how quickly and effectively human drivers understand AV behaviour in complex traffic scenarios. It provides insights into the clarity and predictability of the AV’s actions, and the effort required by human drivers to process and react to AV maneuvers.
While SAGAT remains a widely accepted method for assessing SA, its invasive nature may be unsuitable for continuous tasks like driving. Eye-tracking, especially in 3D VR simulation environments, offers a less intrusive and more practical alternative, ensuring uninterrupted evaluations of driver behaviour.

\subsubsection{Workload} \label{framework_workload}
Workload is broadly defined as the cost of accomplishing task requirements for the human element in human-machine systems \cite{booher2012manprint}. It encompasses cognitive, physical, and temporal demands placed on the operator. Workload can be influenced by a variety of factors, including the number and complexity of tasks or subtasks, time constraints for task completion, the quality of equipment and the working environment, operator skills, strategies, experience, and perception.
Self-assessed workload reported through questionnaires is a common way to empirically measure the operator’s workload. One of the most common is the NASA-Task Load IndeX (NASA-TLX) \cite{steinfeld2006common}. There are also online methods that measure workload through Psychophysiology. The most common ones are: cardiovascular (e.g., heart rate, heart rate variability), brain activity (EEG), eye movement (e.g., variance of eye-gaze coordinates), pupil diameter, skin conductance, respiratory rate, and blink rate. Workload can also be measured through the accomplishment of secondary tasks, where the performance in the second task indicates the residual workload.
Workload is a critical metric for evaluating the difference between HDV-HDV and AV-HDV interactions. Fully autonomous vehicles promise to reduce drivers’ workload, enabling them to focus on secondary tasks. However, this should not come at the expense of increasing the workload for human drivers interacting with AVs.
While subjective methods like NASA-TLX are effective for initial assessments, physiological methods such as eye tracking, especially pupil dilatation \cite{de2014effects}, are less biased and provide real-time insights.

\subsubsection{Smoothness}
Smoothness refers to both the geometry of the trajectory and the velocity and acceleration profiles of AV \cite{kruse2013human, gao2022evaluation}. It plays a critical role in evaluating how fluid and predictable the AV's movements appear, which significantly impacts the perception and comfort of human drivers interacting with the AV.
Two other important parameters are: Path Irregularity (PI) \cite{gao2022evaluation} and Topological Complexity (TC) \cite{gao2022evaluation}.
PI measures the amount of unnecessary turning throughout the AV’s trajectory. A trajectory with high irregularity may appear erratic, reducing smoothness and potentially causing discomfort for human drivers.
TC assesses the level of entanglement in agents' paths. High TC indicates frequent encounters between agents, increasing the likelihood of forced movement adjustments or potential conflicts. Low TC often correlates with smoother and more legible trajectories, facilitating better interactions.
Another way to measure smoothness is through Near-Minimum Jerk \cite{kruse2013human}, which evaluates the extent to which the AV’s motion aligns with the principle of energy optimization observed in human motion. Human motion tends to follow near-minimum jerk trajectories, making such motions appear natural and intuitive. An AV adhering to this principle is more likely to be perceived as smooth and human-like.
Path Irregularity and Near-Minimum Jerk can be calculated by comparing the AV’s trajectory to an ideal path model, providing a quantitative assessment of smoothness. Topological Complexity can be assessed by analyzing the interactions and overlaps between the AV’s trajectory and other agents’ paths.

Smoothness is a non-invasive metric that reflects how human drivers perceive and value an AV’s paths. By evaluating smoothness, researchers can ensure that AV trajectories are not only technically optimized but also intuitive and comfortable for human drivers to interact with. This is particularly important for fostering predictable and cooperative interactions in mixed-traffic environments.\hfill \break

The Interaction Effort domain focuses on reducing the cognitive and physical demands placed on human drivers during AV interactions. Metrics in this domain assess whether AVs are designed and programmed to minimize stress and effort for human drivers, thereby enabling smoother and more efficient interactions.

By identifying sources of high effort or stress, this domain provides actionable insights for improving AV behaviour, ultimately enhancing user experience and safety.

\subsection{DOMAIN 4: Interaction Ability}
The Interaction Ability domain evaluates the AV's capability to perceive, adapt, and respond effectively during interactions with HDVs. This domain emphasizes the AV’s intrinsic qualities, such as self-awareness, human awareness, and autonomy, which are critical for ensuring safe, efficient, and cooperative behaviour in mixed-traffic environments.

\subsubsection{Movement Similarity}
Movement Similarity evaluates the degree to which an AV’s motion resembles human motion, focusing on trajectory alignment and motion behaviour. This metric provides insights into how naturally and intuitively the AV interacts with human drivers by comparing its trajectory to human-driven vehicles \cite{kruse2013human, gao2022evaluation}.
A key metric for assessing Movement Similarity is Displacement Errors.
One can find different ways to assess Displacement Errors in literature, namely through:
\begin{itemize}
    \item Average Displacement Error (ADE): Measures the average distance between the AV’s trajectory and the human trajectory over the entire path
    \item Final Displacement Error (FDE): Represents the distance between the final destinations of the AV and human trajectories.
\end{itemize}
ADE and FDE are often used together to provide a comprehensive comparison of spatial alignment.

Another way to measure Movement Similarity is through Dynamic Time Warping Distance (DTWD), which evaluates differences in motion behaviours at varying speeds by aligning trajectories temporally, and 
a modified DTWD. This variant uses time re-scaling to transform one trajectory into another, allowing for a more nuanced comparison of dynamic behaviours.
DTWD is particularly useful for capturing differences in velocity and timing between AV and human motions \cite{gao2022evaluation}.

Lastly, Path Compatibility, introduced by Kruse et al. \cite{kruse2012legible}, assesses whether two agents’ paths can be followed concurrently without causing deadlocks, ensuring that both agents reach their destinations with minimal deviation. Higher path compatibility reduces the planning and cognitive effort required by both agents, leading to smoother and more cooperative interactions.

Movement Similarity metrics require human trajectory data to establish a baseline for comparison. This can be achieved through experiments where two human drivers interact in a controlled scenario to generate reference trajectories or where the same experiment is repeated with one HDV and one AV, allowing for direct comparisons between their trajectories using ADE, FDE, and DTWD.

Movement Similarity highlights how closely the AV mimics natural human motion, which is crucial for fostering trust and predictability in interactions. By ensuring high similarity, AVs can create more intuitive and cooperative dynamics with human drivers, reducing confusion and enhancing safety in mixed-traffic environments.

\subsubsection{Quantitative Performance}
Quantitative Performance \cite{young2011evaluating} evaluates the \textbf{effectiveness} and \textbf{efficiency} of the interaction between an AV and other agents. It focuses on measurable outcomes that reflect the AV’s ability to achieve task objectives and perform optimally in various scenarios.

\textbf{Effectiveness} also referred to as Task Completion \cite{young2011evaluating}, is defined as the percentage of tasks successfully performed during the interaction. It measures whether the AV achieves its intended goals, such as completing a merging manoeuvre.
\textbf{Efficiency} is defined as the time required to perform the task(s) successfully. It captures how quickly and smoothly the AV completes its objectives, reflecting operational optimization.

Quantitative performance is typically analyzed using:
\begin{itemize}
    \item Success rates: Evaluates the frequency of task completion.
    \item Velocity and acceleration profiles: Assesses the smoothness and consistency of movement.
    \item Statistical analysis: Metrics like averages and standard deviations are used to summarize performance over multiple trials.
\end{itemize}

Quantitative performance metrics are foundational to AV evaluation and are widely used in assessing AV controller algorithms. As highlighted earlier, metrics such as success rates for merging manoeuvres and velocity profiles are already embedded in AV evaluations.

\subsubsection{Awareness of Humans}
Human Awareness refers to the degree to which an AV is “aware” of human drivers and their actions \cite{steinfeld2006common, drury2003awareness}. This involves the AV’s ability to detect, track, and adapt to human behaviour, ensuring safe and effective interactions. Human awareness is critical for fostering intuitive and cooperative dynamics in mixed-traffic environments.

Key components of Human Awareness are:
\begin{itemize}
    \item Human Detection and Tracking: The AV’s ability to identify and monitor human drivers and their vehicles in real time. Accurate detection and tracking are essential for anticipating human actions and avoiding collisions.
    \item User Modelling and Monitoring: Understanding and predicting the behaviour of human drivers based on observed data. This enables the AV to tailor its responses to individual drivers, enhancing interaction quality.
    \item Adapting to User Behaviour: The AV’s ability to adjust its actions dynamically in response to human drivers’ behaviour. It supports smooth and cooperative interactions, aiming to reduce the cognitive effort required from human drivers.
    \item Awareness Violation: Situations where the AV fails to provide expected awareness or responses, leading to misinterpretations by human drivers. Minimizing awareness violations is crucial for maintaining trust and safety during interactions.
\end{itemize}

The level of required human awareness is directly proportional to the AV’s level of autonomy \cite{steinfeld2006common, drury2003awareness}. Higher autonomy demands greater sensitivity to human behaviour and context to ensure seamless integration into mixed-traffic systems.
Human awareness is a foundational metric for assessing how well an AV interacts with human drivers. By detecting, modelling, and adapting to human behaviour, the AV can anticipate potential conflicts, reduce misunderstandings, and foster trust. Incorporating human awareness into AV evaluation ensures the system aligns with human expectations, promoting safer and more intuitive interactions.

\subsubsection{Autonomy}
Autonomy refers to the degree to which an AV can perform tasks independently without requiring human intervention. A key metric for evaluating autonomy is neglect tolerance \cite{scholtz2003theory}, introduced by Goodrich et al. \cite{goodrich2003seven}. Neglect Tolerance measures how the AV’s task effectiveness declines over time when neglected by the user. It highlights the resilience of the AV to limited human input, ensuring safe and effective performance even when the operator fails to react or interact with the system. If effectiveness declines below a specific threshold, the AV’s autonomy becomes insufficient for practical use.

Neglect tolerance can be affected by different factors like time delays in remote operations, task complexity, robot capability, user interface, and the user \cite{scholtz2003theory, goodrich2003seven}.
Autonomy metrics, including neglect tolerance, are already widely used to evaluate AV controllers. However, autonomy remains a foundational aspect of AV evaluation, ensuring the system can function effectively with minimal human intervention.

\subsubsection{Self-Awareness}
Self-Awareness refers to the degree to which an AV can accurately assess its state, capabilities, and limitations \cite{steinfeld2006common}. This characteristic directly influences the AV’s ability to cooperate effectively with human drivers, ensuring smoother and more predictable interactions.
Steinfeld et al. \cite{steinfeld2006common} propose the following aspects as essential for assessing self-awareness in robots:
\begin{itemize}
    \item Robots Limitations: The AV’s ability to recognize the constraints of its sensors, actuators, and other systems. It enables the AV to operate safely within its functional boundaries.
    \item Self-Monitoring: Continuous monitoring of critical parameters to detect deviations or anomalies. It ensures real-time awareness of the AV’s operational state.
    \item Fault Detection and Resolution: The ability to identify faults and autonomously implement corrective actions. It minimizes reliance on external intervention and prevents potential safety risks.
\end{itemize}
Self-awareness is a qualitative metric that highlights an AV’s capability to maintain operational integrity. By understanding its limitations, monitoring critical parameters, and resolving faults autonomously, the AV enhances its reliability and supports effective cooperation with human drivers.

Incorporating self-awareness metrics into AV evaluation ensures that the system is not only functional but also resilient and capable of adapting to dynamic traffic scenarios.
\hfill \break

The Interaction Ability domain focuses on the AV’s capacity to function effectively and harmoniously in complex, dynamic environments. By evaluating self-awareness, human awareness, autonomy, and movement similarity, this domain ensures that the AV is not only capable of executing tasks but also of interacting seamlessly with human drivers.

This domain bridges the gap between technical performance and interaction quality, emphasizing the AV’s ability to anticipate and adapt to human behaviour while maintaining safe and efficient operation. Together, these metrics provide a holistic understanding of the AV’s competence in real-world traffic scenarios.

\section{Case Study}
\begin{figure*}
    \centering
    \captionsetup{justification=raggedright}
    \includegraphics[width=1\textwidth]{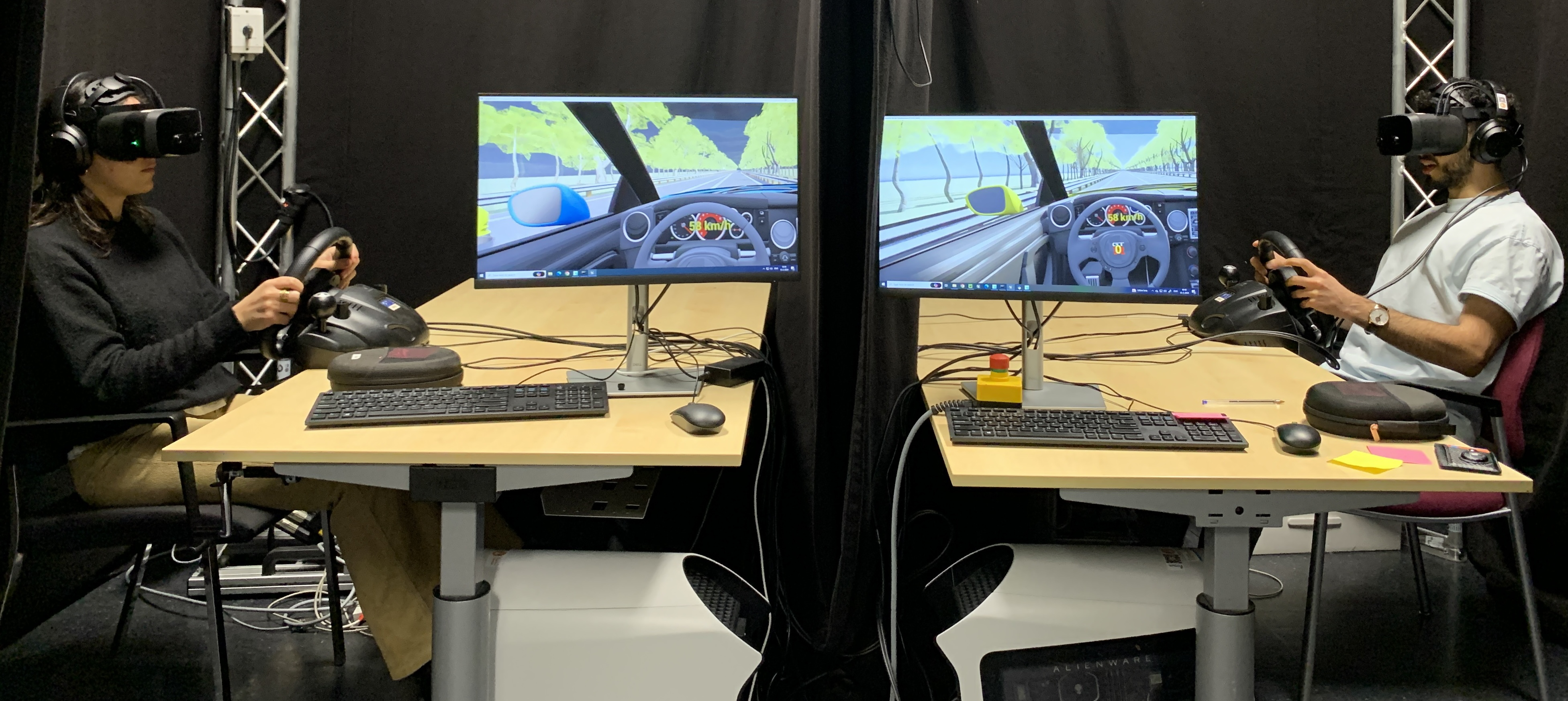}
    \caption{Experimental setup. Two participants taking part in the experiment each with a steering wheel, pedals and VR headsets}
    \label{setup}
\end{figure*}

To demonstrate the usefulness of our framework, we performed a case study in which we evaluated a state-of-the-art interaction-aware controller for autonomous vehicles~\cite {sadigh2018planning}. In our case study, we compare merging interactions between two human drivers with interactions between a human and an AV. Our framework is used to select the additional metrics and methods to evaluate the controller (besides the traditional metrics for safety and efficiency). We conducted our case study in a coupled VR-based driving simulator (\autoref{setup}). In this simulator, two participants can interact and look around while driving separate vehicles. The VR setup allowed for measuring many different signals, making it the ideal setup to apply the framework.

\subsection{Assessment of human-robot interactions; applying the framework}
The framework (\autoref{framework}) maps the full range of important aspects in interactions between humans and AVs. We used its four domains to ensure our evaluation in this case study reflects that depth in interaction aspects. Specifically, we used the framework as a guide to select one metric per interaction domain. This keeps the evaluation tractable while covering the important aspects of an interaction.

\subsubsection{Interaction Effect - Psychological Safety}
The interaction effect domain for merging interactions is limited since physical and spatial safety are directly related to collision prevention, which is covered in the traditional evaluation of AVs. Trust in technology is not relevant for most merging interactions since drivers do not know if they are interacting with a human or an AV. Therefore, we measured psychological safety to evaluate the interaction effect domain.

Psychological safety encompasses perceived safety and stress levels. Typically, questionnaires are employed to assess the user's perceived psychological safety (e.g.,~\cite{bartneck2009measurement, carpinella2017robotic, joosse2013behave}), focusing on aspects such as the absence of obstructions, the ability to maintain preferred velocity in the robot's presence, and overall impressions of the interaction. Other methods include using biometrics such as heart rate (variability), pupil dilation, or gaze direction~\cite{rubagotti2022perceived}. However, we used a different approach: self-reported perceived safety using a direct input method~\cite{rubagotti2022perceived}. This entails that participants continuously report their perceived safety using steering wheel buttons. This is a simple, economical, and non-invasive method to measure perceived safety continuously throughout the interaction.

\subsubsection{Interaction Perception - Subjective Ratings}
We used subjective ratings from the interaction perception domain to investigate whether participants could distinguish between interacting with human-driven vehicles and AVs. The expectation was that if participants could independently notice differences in driving behaviours without prior knowledge of the type of vehicle they were interacting with, it would suggest that there are observable and distinguishable behavioural patterns associated with AVs compared to HDVs. 

To investigate if participants could make this distinction, we told them they were constantly interacting with the other participant. We did not inform participants that they would encounter an autonomous vehicle in half of the trials. Human and autonomous vehicles had different colours (which also were not mentioned in the instructions) to enable participants to report on them separately. In instances where a vehicle exhibited particularly notable behaviour, the colour association was intended to aid participants in recalling and linking these behaviours to specific vehicles during the post-experiment questionnaires. While if the driving behaviour of the interacting vehicles would not differ notably, the colour difference was probably disregarded.

After the trials, each participant was asked to fill in a questionnaire to assess if they noticed any difference in behaviour between the colour-coded vehicles. The following questions were asked:
\begin{enumerate}
    \item To which extent were you able to understand the other participant’s intentions? (on a scale 1-5)
    \item Have you noticed any differences in behaviour between cars of different colours? (Yes/No)\\
    \textbf{\textit{If Yes:}}
    \item How large were which differences in behavior that you noticed? (on a scale 1-5)
    \item How exactly did the behaviour of the other cars differ depending on their colour? (open-ended)
    \item How did the different behaviours of the other cars influence your interaction with them? (open-ended)
\end{enumerate}

\subsubsection{Interaction Effort - Workload}
For interaction effort, we selected the workload metric, a key parameter describing the cost of task completion within human-machine systems~\cite{booher2012manprint}. Various factors, including task complexity, time constraints, equipment quality, working conditions, task performance, operator skills, strategies, experience, and perception can influence workload. One of the key advantages of AVs is their ability to significantly reduce the workload of their users, allowing them to focus on other tasks such as working or relaxing. However, this may come at the cost of increasing the cognitive workload of other road users, depending on the AV's ability to interact without requiring (additional) effort. Therefore, it is crucial to evaluate how AV design influences the workload of human drivers.

Workload can be measured empirically in multiple different ways, e.g., using self-assessment~\cite{steinfeld2006common}, secondary tasks~\cite{ogden1979measurement}, or by measuring biological signals such as cardiovascular indicators~\cite{jorna1993heart,jorna1993heart}, brain~\cite{brouwer2012estimating} or respiratory~\cite{veltman1998physiological} activity, or eye-tracking~\cite{de2014effects,liu2022assessing, nilsson2020off}. We chose the last option since eye-tracking provides a non-intrusive, cost-effective way to assess workload. Furthermore, eye-tracking can easily be used in VR because some head-mounted displays offer integrated pupil tracking. 

\begin{figure}[h]
    \centering
    \captionsetup{justification=raggedright}
    \includegraphics[width=8.7cm]{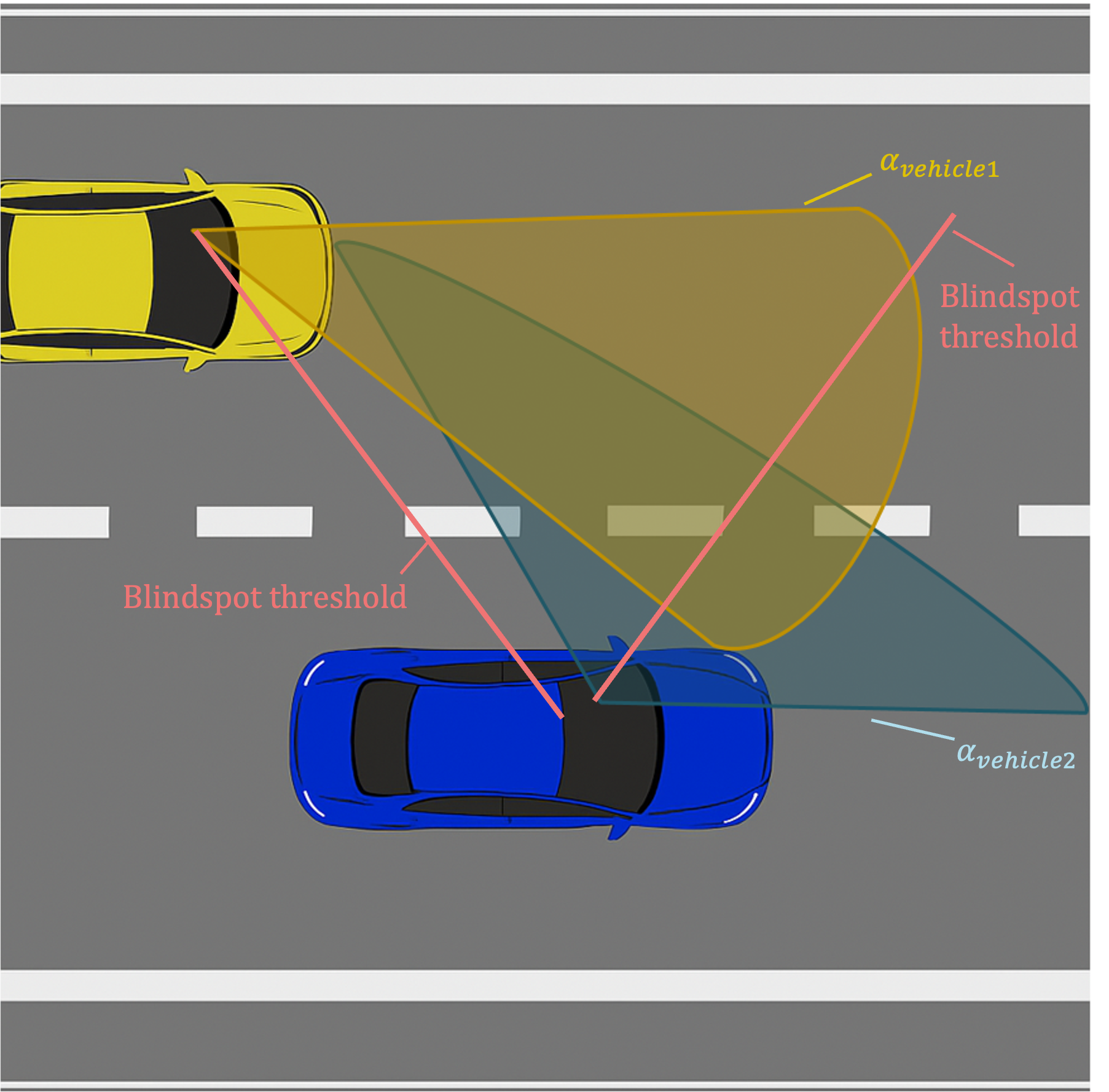}
    \caption{The angle of the two vehicles is computed by taking the driver's position inside of their vehicle to the first part of the other vehicle where the driver can see if the vehicles are parallel. For the car on the highway lane (yellow car in this example), the angle $\alpha_{vehicle1}$ is between the driver's position and the front left part of the blue vehicle. Conversely, for the vehicle in the on-ramp, the angle $\alpha_{vehicle2}$ is computed between the driver's position and the most front-right part of the other vehicle. In addition to the angles $\alpha_{vehicle1}$ and $\alpha_{vehicle2}$ a blind spot threshold for the driver's gaze is introduced. This threshold has an absolute value of 50º.}
    \label{angleVE}
\end{figure}

We use the driver's gaze angle (the sum of head rotation angle and eye movement angle: \(\alpha_{gaze} = \alpha_{head} + \alpha_{eye}\)) to determine workload under the assumption that the frequency and duration of looking at the other vehicle in an interaction increases with workload~\cite{liu2022assessing}. Whenever the absolute value of $\alpha_{gaze}$ or its foveal region exceeded the angle between the two vehicles ($\alpha_{vehicle1}$ (\autoref{angleVE}) for Driver 1 and $\alpha_{vehicle2}$ for Driver 2) or exceeded the blind spot threshold (a fixed threshold of 50 degrees), it was interpreted as the driver verifying the presence of the other vehicle.

\subsubsection{Interaction Ability - Quantitative Performance}
To evaluate the ability of the AV to interact, we compared quantitative performance in terms of high-level outcomes of the interaction between the human-human and human-AV trials. We chose this metric because we are interested in comparing the AV's ability to interact directly with an interaction between two human drivers. Comparing quantitative performance is the easiest method to do this. Specifically, we evaluated the number of collisions and the percentage of trials where the merging vehicle merges in front of the other vehicle. 

\subsection{Methods} \label{Expdesign}
In our experiment, two participants drove in a merging scenario in a virtual reality-based driving simulator. Our simulator uses Unreal Engine 4.26, CARLA 0.9.13~\cite{Dosovitskiy17}, and JOAN: an open-source software framework~\cite{beckers2023joan}. Participants used a USB steering wheel with pedals to control their vehicle (Logitech Driving Force GT). Two buttons on the steering wheel were used to report perceived safety. We used two Varjo VR3 VR headsets, which have built-in eye-tracking.

Participants were instructed to adhere to their typical driving behaviour. Verbal communication was expressly prohibited; participants wore noise-cancelling headphones to enforce this. Finally. Participants were informed they would only interact with each other, with no mention of interactions with AVs. This served three purposes. First, to replicate a real-world scenario where drivers might not always know if they are interacting with an AV or an HDV. Second, to prevent potential biases (positive or negative) 
in participants' attitudes or beliefs about AVs to influence the results. Third, to prevent the participants from actively seeking out differences between AVs and HDVs, which could lead to exaggerated perceptions of differences that, under normal traffic conditions, might not be noticeable or significant.

\begin{figure}[h]
    \centering
    \captionsetup{justification=raggedright}
    \includegraphics[width=8.7cm]
    {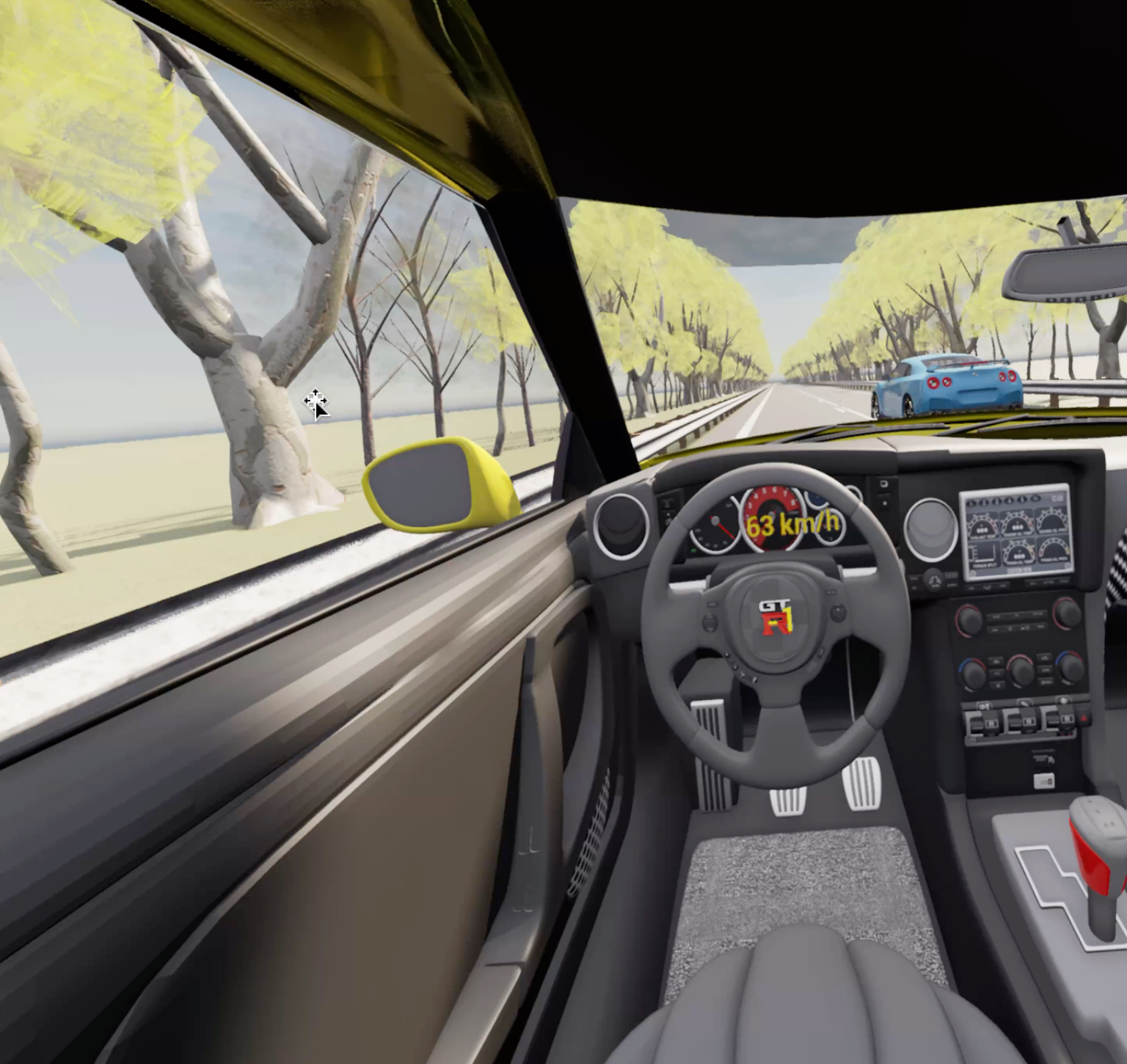}
    \caption{Participants' Point of View inside of their vehicle during the experiment. In the middle of the steering wheel the participant can see the value of their Perceived Safety}
    \label{POV}
\end{figure}

\begin{figure*}[h]
    \centering
    \captionsetup{justification=raggedright}
    \includegraphics[width=1\textwidth, height=4cm]{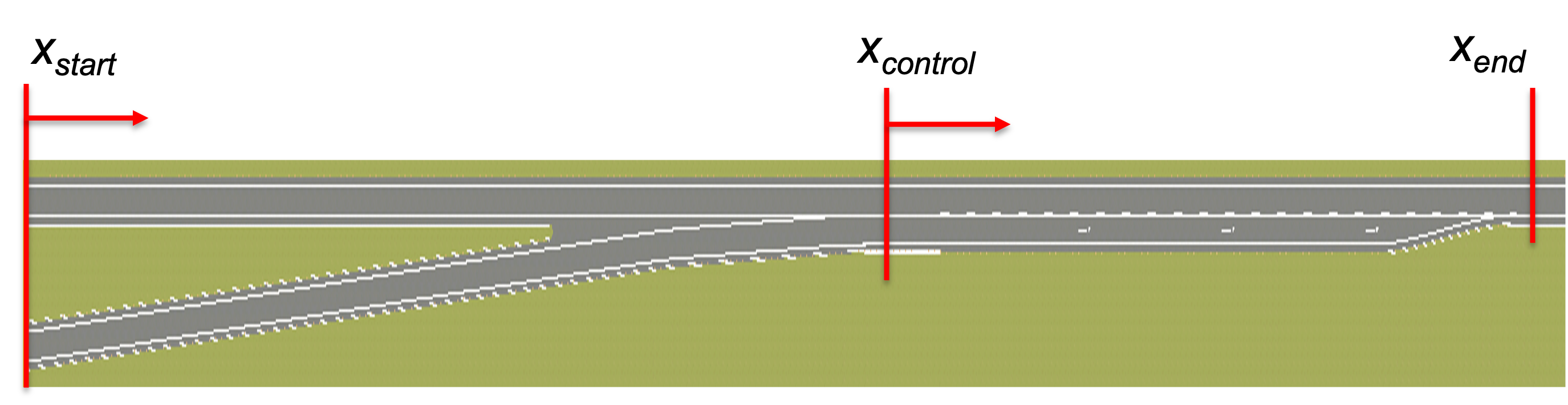}
    \caption{Track used for the experiment, with merging lane and highway lane. Where $x_{start}$ shows the starting positions, $x_{control}$ shows where the participants get control and $x_{end}$ shows the end of the merging lane}
    \label{Track}
\end{figure*}

Participants experienced a first-person perspective (\autoref{POV}), driving one-lane road with an additional merging lane (\autoref{Track}). Initially, the vehicles were operating under cruise control, where the participant could only steer the vehicle. This was done to have precise control over the kinematic conditions at the merge~\cite{Siebinga_2024, Siebinga_2024b}. At the beginning of the merging lane (denoted $x_{control}$ in \autoref{Track}), at the same time for both drivers, participants assumed full control. This was indicated with road signage and accompanied by an auditory cue from the headphones. The trial ended when the first participant reached the end of the road ($x_{end}$).

We used four different experimental conditions (\autoref{table_conditions}). In two of these, the participants were driving on either the highway lane or on-ramp and interacted with an AV (AV-HDV). In the other two conditions, the participants interacted with each other; these were independently evaluated and labelled according to the driver of interest (highway with human/on-ramp with human).

Participants drove 10 training trials against each other, after which each experimental condition was repeated ten times in a randomized order, yielding a total of 40 trials per participant. The experiment's speed limit was 80 km/h; participants were explicitly instructed only to exceed this limit under critical circumstances.

Each vehicle was designated by a specific colour code (yellow, blue, green, and grey). Yellow and blue corresponded to the vehicles operated by the participants, with participant 1 assigned to the yellow car and participant 2 assigned to the blue car. The AVs' colours were green and grey. Participants interacted with all interaction partners, thus with three different coloured-coded vehicles. After the experimental sessions, participants were asked to complete a questionnaire to describe any behavioural distinctions between the different coloured vehicles.

\begin{table}
    \caption{Conditions and interaction type from the perspective of Driver 1}
    \label{table_conditions}
    \begin{tabular}{c|c|c|c|c}
    \hline
    Condition     & Driver 1 & Driver 2 & AVs  &Interaction\\ \hline
    Highway with & Highway & On- & \multirow{2}{*}{-} & \multirow{2}{*}{HDV-HDV} \\ Human & lane & Ramp & & \\ \hline
    On-Ramp with & On- & On- & Highway & \multirow{2}{*}{AV-HDV} \\ AV & Ramp & Ramp & lane & \\ \hline
    On-Ramp with & On- & Highway & \multirow{2}{*}{-} & \multirow{2}{*}{HDV-HDV} \\ Human & Ramp & lane & & \\ \hline
    Highway with & Highway & Highway & On- & \multirow{2}{*}{AV-HDV} \\ AV & lane & lane & Ramp & \\ \hline
    \end{tabular}
\end{table}

\subsection{AV Controller}
A state-of-the-art interaction-aware controller~ \cite{sadigh2018planning} controlled the AV, to which we made slight modifications to make it suitable for our setup. Specifically, we replaced the point-mass dynamics model with a bicycle model to allow integration in Unreal Engine. Furthermore, we adjusted the maximum acceleration and deceleration values alongside the maximum steering angle to replicate those of the vehicle in Carla. Because of these changes in dynamics and scenario, we needed to re-tune the values of the reward function weights. The AV planner ran on a separate Ubuntu computer connected to the simulator server via a network connection.

Twenty participants (6 female/14 male, mean age 25.7, sd 4) took part in our experiment, they were randomly assigned to ten pairs. All participants possessed a valid driving license, and none wore glasses or contact lenses (required for the eye-tracking VR). All the participants gave informed consent and were compensated with a €15 gift card. Most of the participants were students and researchers of TU Delft, thus already familiar with the concept of autonomous vehicles, autonomous vehicles' algorithms and the field of Human-Robot Interaction. The experiment received ethical approval from the ethical committee of TU Delft.

\subsection{Results}
We analysed the data from the moment the drivers took control until the merging manoeuvre was completed (meaning the merging vehicle crossed the line marking) since we were interested in the interaction itself. We compared the condition "Highway with human" to "Highway with AV" and the condition "On-Ramp with human" to "On-Ramp with AV". Thereby evaluating the workload and perceived safety associated with the same manoeuvres (merging or staying in the lane) in HDV-HDV and AV-HDV interactions.

\subsubsection{Interaction Ability - Quantitative Performance}
Each experimental trial had three possible high-level outcomes: the merger merges ahead or behind the highway vehicle, or they collide. In total, 14 collisions occurred. Collisions were less frequent in human-human interactions compared to human-AV interactions, yet this difference is not statistically significant. 

These collisions were excluded from the further analysis for two reasons. Firstly, our primary research interest is understanding typical AV-HDV merging interactions. Collisions represent extreme cases, thereby not providing useful insights into the normal range of behaviours we aim to study. Secondly, including collision data could introduce significant outliers and thus affect the results, leading to misinterpretations about typical driver behaviour and interaction patterns. 

In total, 200 trials were conducted per condition. The collision rates were 1\% for both Highway with human (2 collisions) and On-Ramp with human (1 collision), and 3\% for both Highway with AV (5 collisions) and On-Ramp with AV (6 collisions).

\begin{figure}[h]
    \centering
    \captionsetup{justification=raggedright}
    \includegraphics[width=8.7cm, height=6cm]{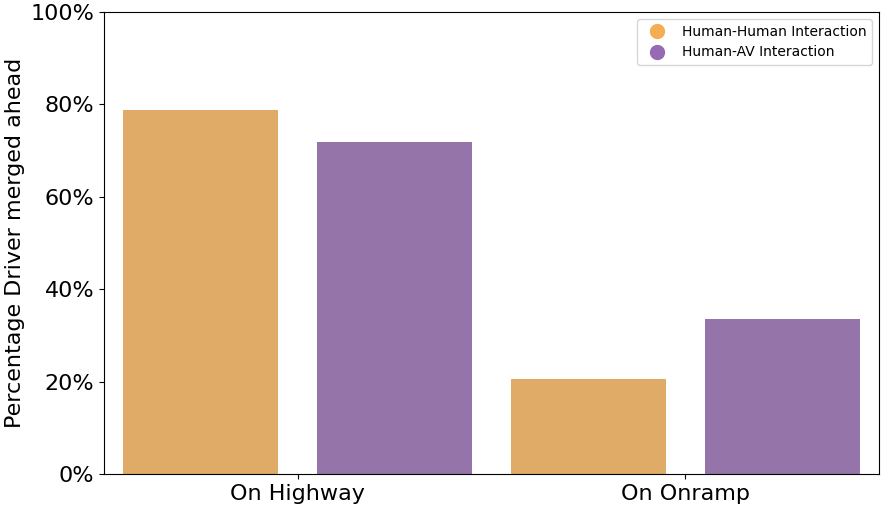}
    \caption{Percentage of merges in front of the other vehicle per condition}
    \label{barplotcountmerges}
\end{figure}

In terms of a driver going first or not, the vehicle already on the highway stays in front most of the time and lets the merging vehicle merge behind (\autoref{barplotcountmerges}). The difference between interaction between humans and a human and an AV is insignificant. In "On-Ramp" scenarios, the merging vehicle tends to merge in front less often. This difference between interaction partners is statistically significant (\(p = 0,024\)).

\subsubsection{Interaction Effort - Workload}
\begin{figure*}[h]
    \centering
    \captionsetup{justification=raggedright}     \includegraphics[width=\textwidth]{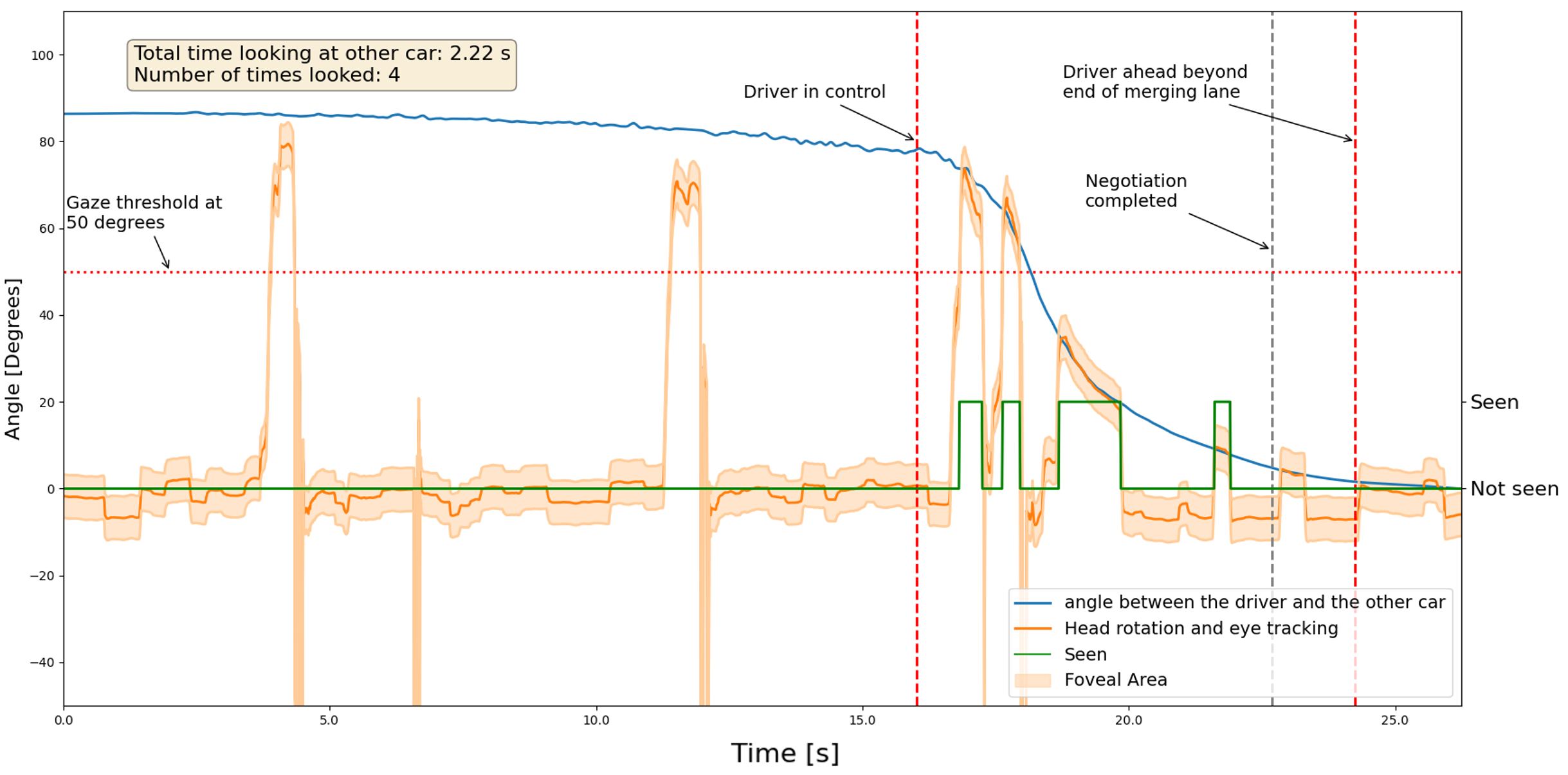}
    \caption{Eye-tracking data of a representative trial with two human drivers where the ego driver is in the highway lane. The orange line shows the gaze angle and foveal region, the blue line represents the angle between the vehicles. We automatically detected fixations on the other vehicle by denoting when the driver's gaze angle or its focal region exceeds the threshold ($50^\circ$) or $\alpha_{vehicle1}$ (blue line). In this trial, the total fixation time was 2,22s with 4 fixations. The driver sometimes looks directly at the other vehicle, but in some cases, they only check its presence by looking roughly in the other vehicle's direction. This is mainly done when the angle between the vehicles is large and this behaviour was the reason for introducing the threshold in fixation detection.}
    \label{eyedata}
\end{figure*}

We used eye-tracking to measure workload, assuming that an increase in duration of fixations on the other vehicle indicates a higher workload. \autoref{eyedata} shows a representative trial with two human drivers. We analysed the data in the time window from when the drivers gain control until the end of negotiation (the moment when the merging vehicle's centre crosses the merge lane marking).

\begin{figure}[h]
    \centering
    \captionsetup{justification=raggedright}
    \includegraphics[width=8.7cm, height=7cm]{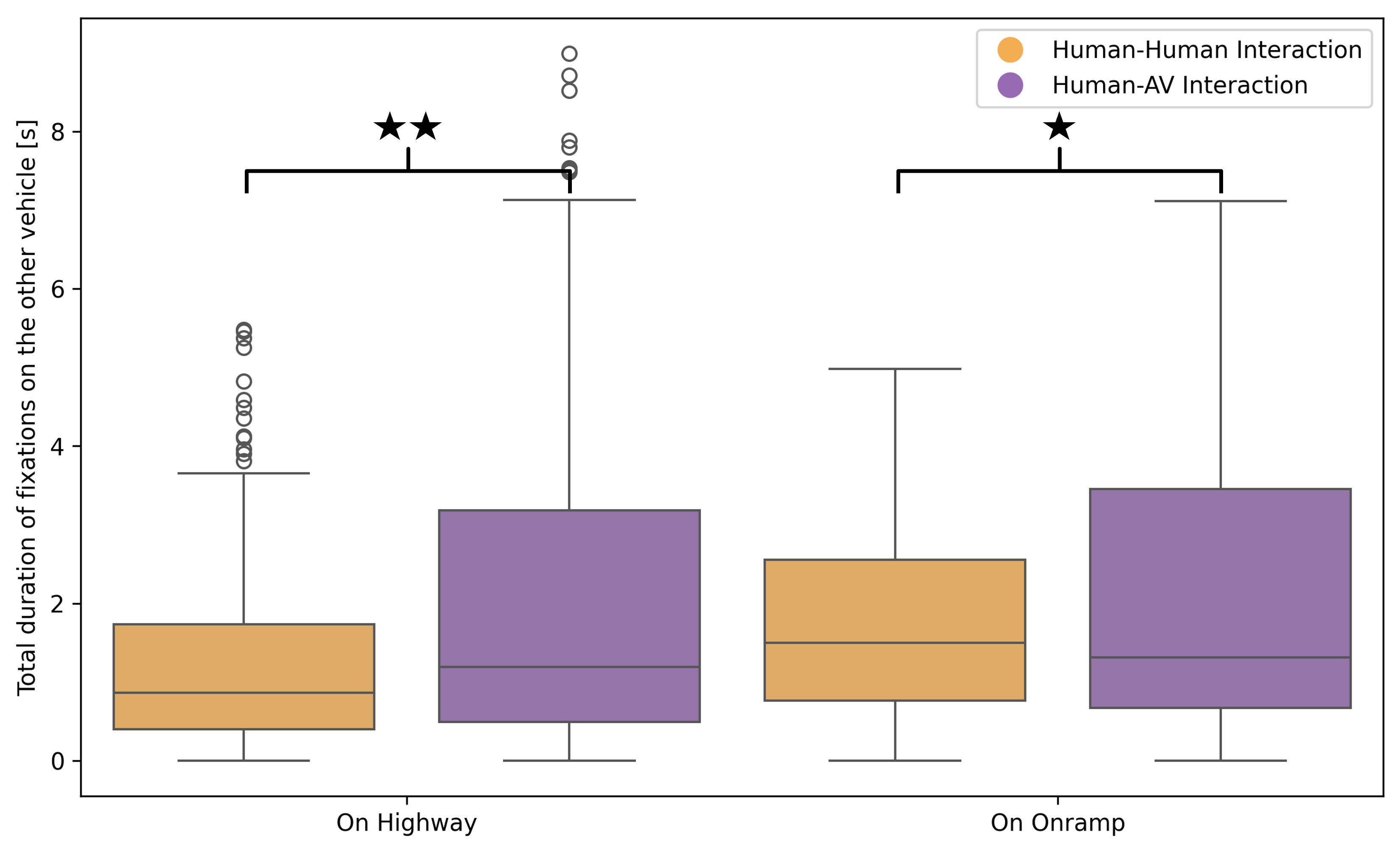}
    \caption{Total duration of fixations on the other vehicle per experimental condition.}
    \label{boxplottime}
\end{figure}

The total time spent looking at the other vehicle was significantly higher for interactions with the AV in both lanes(\autoref{boxplottime}). This confirms the hypothesis that drivers need more time to understand the AV's behaviour in a merging scenario.

\subsubsection{Interaction Effect - Perceived Safety} \label{PSR}
\begin{figure}[h]
    \centering
    \captionsetup{justification=raggedright}    \includegraphics[width=8.7cm, height=6cm]{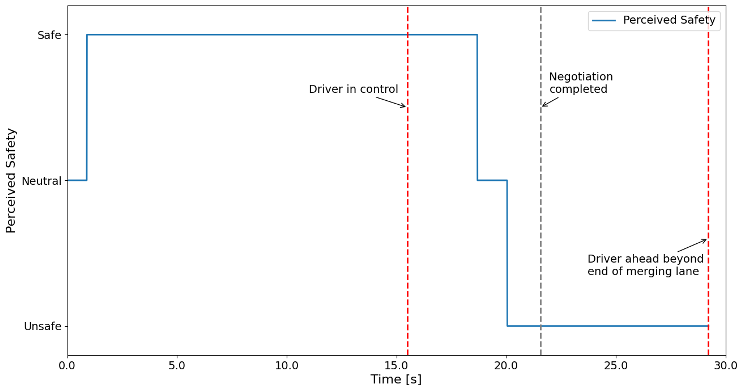}
    \caption{Perceived Safety for a representative trial (’On-Ramp with AV’ condition). The blue line represents the Perceived Safety signal; the dashed red lines indicate the moments when drivers gain control and when the merging vehicle reaches the end of the merge lane; the dashed grey line indicates completion of the merging manoeuvre.}
    \label{riskrow}
\end{figure}

\begin{figure}[h]
    \centering
    \captionsetup{justification=raggedright}
    \includegraphics[width=8.7cm, height=7cm]{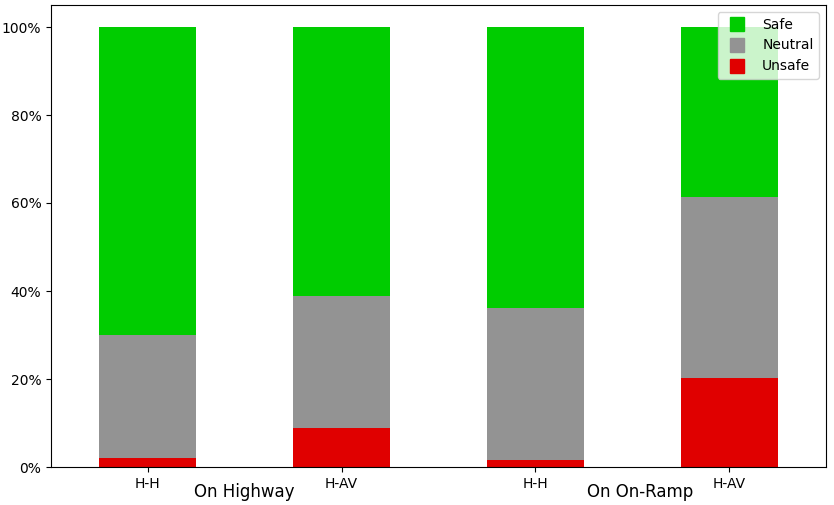}
    \caption{Perceived safety when the negotiation ended}
    \label{boxplotsafety}
\end{figure}

\begin{figure*}[h]
    \centering
    \captionsetup{justification=raggedright}
    \includegraphics[width=0.99\linewidth]{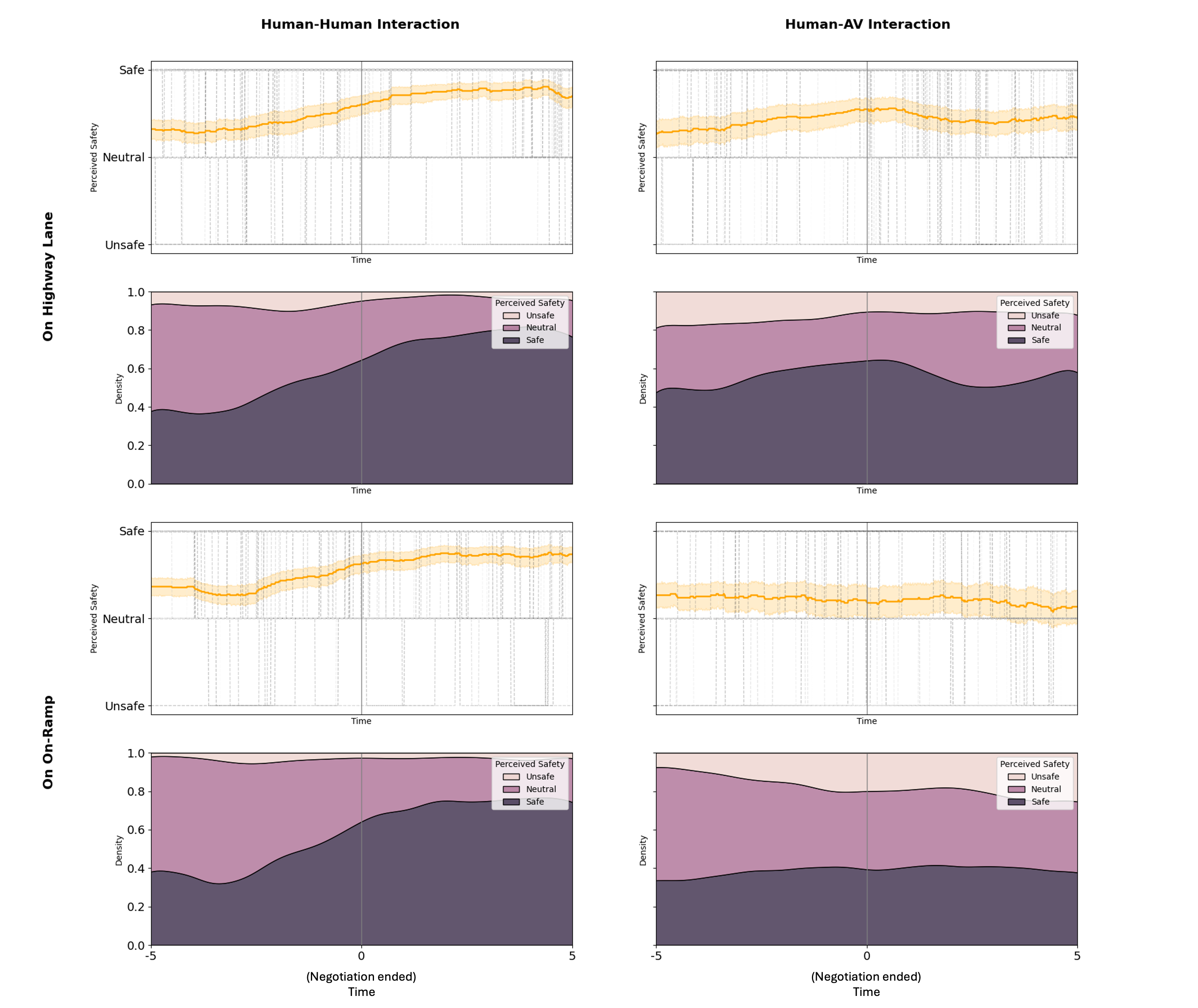}
    \caption{Aggregated Perceived Safety per condition. The left panels show human interactions and the right ones show AV interaction. Each dotted line represents an individual trial; the continuous orange line and surrounding shaded area illustrate the mean and the standard deviation. The vertical grey dotted line marks the end of the negotiation, the trials are aligned around this moment.
    }
    \label{PS_all}
\end{figure*}

Perceived Safety was measured continuously throughout the experimental trials. \autoref{riskrow} shows a representative example with a shift in the driver's perceived safety: a decrease when the merging interaction started. The interaction's impact on the driver's perceived safety can be measured particularly at the end of the negotiation: at the moment when the merging vehicle crosses the lane marking. The aggregated results at that moment in the trial show that drivers generally felt safe less often after interacting with the AV (\autoref{boxplotsafety}). Also, participants felt unsafe more often after these interactions. Aggregated time traces around this moment when the interaction is finished show the same trend (\autoref{PS_all}), yet we did not perform statistical analysis on these time signals.

\subsubsection{Interaction Perception - Subjective Ratings}
The questionnaires aimed to measure if drivers perceived behavioural differences between colour-coded vehicles. Participants generally rated their understanding of other drivers' intentions as neutral, with an average score of 3.5 and a standard deviation of 0.69. A majority (65\%, 13/20) observed behavioural variances between different coloured cars, with 61.6\% considering these differences substantial (average rating of 3.75 and a standard deviation of 1.22).

In the following answers, the green and grey vehicles were AVs, and the blue and yellow vehicles were HDVs. 

Participants were asked: \textit{“How exactly did the behaviour of the other cars differ depending on its colour?”}. Although they were not informed that the green and grey cars were AV and the blue and yellow cars were HDV, many still perceived behavioural differences between vehicles, often associating them with specific colours.
A number of participants described the AVs (green and grey vehicles) as behaving in ways that felt less predictable or more difficult to interpret. Comments such as \textit{“The green car} [AV] \textit{didn’t know what he wanted to do — accelerate or stop”} (participant 2), and \textit{“abrupt speed changes, non-predictable driving behaviour”} (participant 15) reflected a general uncertainty about the other vehicle’s intentions.
Some participants also noted differences in smoothness, positioning, and acceleration, with remarks like \textit{“The grey car} [AV] \textit{did not seem to accelerate smoothly”} (participant 12) or \textit{“The green and grey cars} [AVs] \textit{would slow down, skid and crash”} (participant 4). Others mentioned that vehicles \textit{“seemed more aggressive”} or \textit{“tried to get ahead”}, with one participant summarizing: \textit{"Yellow} [HDV] \textit{– normal behaviour;} \textit{Green or grey} [AVs] \textit{– aggressive, reckless”} (participant 20).
Even when not explicitly tied to a colour, several responses suggested that certain vehicles felt more assertive, aggressive, or difficult to read — indicating that participants were forming judgments based on perceived driving style, even in the absence of clear vehicle-type information.
Overall, these qualitative responses suggest that participants intuitively recognized differences in driving behaviour, with many of the descriptions aligning with common perceptions of AVs as less legible or natural in traffic interactions. This expressed frustration with the AVs' unclear intentions reinforces the importance of evaluating AV behaviour from a human perception standpoint, as part of a broader understanding of how AVs are experienced in mixed traffic environments.

We divided the answer to the final question \textit{How did the different behaviours of the other cars influence your interaction with them?} in two groups:
\begin{enumerate}
    \item \textit{\textbf{Proactive Participants}}: Those who took the initiative by speeding up: \textit{"Speed up and merge in front of them"} (participant 4),  or overtaking, especially when foreseeing potentially dangerous behaviour from other vehicles: \textit{"when I could foresee a dangerous behaviour because of the colour of the car I would tend to take initiative to overtake it so that its behaviour and speed changes does not affect me"} (participant 15)
    \item \textit{\textbf{Cautious Participants}}: Those who responded more conservatively, maintaining larger distances: \textit{"Larger distance = safer perception"} (participant 5) or waiting for the other vehicle to act first: \textit{"waited for the action of the other participant"} (participant 7), especially when perceiving the behaviour as reckless or aggressive: \textit{"was more cautious when the behaviour seemed more reckless"} (participant 8). Participant 20 summarises: \textit{"Was much more careful around the green and grey cars, let them go ahead"}
\end{enumerate}

Five out of 13 participants reported that they took the initiative, while 7 participants described their reaction as more conservative. None of the participants reported a complete lack of change in their behaviour in response to the change in the behaviour of other vehicles. However, one participant did mention that their behaviour was only slightly influenced, stating: "\textit{Not that much, I was more careful around the grey car}". This near-even decision highlights that drivers use diverse strategies in response to perceived differences in the behaviours of AV vehicles. However, all drivers who noticed a difference in behaviour also responded. 

\subsection{Case-study: Discussion}
In this case study, we applied our proposed framework (\autoref{framework}) to evaluate a state-of-the-art autonomous vehicle controller. We conducted a driving-simulator experiment to assess how the AV performs in terms of interaction effect, interaction perception, interaction effort, and interaction ability. Following the framework, we analysed high-level outcomes, driver workload, perceived safety, and subjective ratings.

We found an increased workload in interaction with the AV, suggesting that this AV requires drivers to allocate more attention to understand its behaviour, potentially leading to increased cognitive demands \cite{liu2022assessing}. Understanding the root causes of this workload increase is critical for improving general AV integration into human-driven traffic. We also found a decreased perceived safety in "On-Ramp with AV" conditions. This could be due to reduced predictability and smoothness of the AV. Which also affects drivers’ reactions to AV behaviour. Although these findings cannot be easily generalized to all AV controllers, they highlight the necessity of evaluating AVs to identify and mitigate any negative aspects of their ability to interact with human drivers, thereby showing the value of our proposed approach.

The answers from participants to our questions about the AV’s predictability and comprehensibility could be linked to the increased workload. As drivers spend more effort to interpret the AV’s behaviour, they experience greater cognitive demands, which may negatively impact their behaviour and perception of the other vehicle. The perceived behaviours of AVs as \textit{"non predictable"}, \textit{"more aggressive"} and \textit{"reckless"}, as described by the participants, could be the cause of the reduced perceived safety in the "On-Ramp with AV" condition. This suggests a link between the participants' subjective experiences and the study's quantitative findings. This indicates that the AV behaviour has an effect on human driver perception and responses, even in short interactions such as merging where human drivers are unaware whether the other vehicle is an AV or human-driven.

The AV successfully interacted with drivers without significantly increasing collision risk. The high-level outcomes, such as the percentage of merges in front of the other vehicle, showed only a slight difference with respect to human-human interactions. When relying solely on these traditional metrics, one would concluded that the AV’s merging ability was comparable to that of human drivers. However, by including interaction metrics, we uncovered substantial differences in driver workload and perceived safety, highlighting potential negative impacts on human drivers. These findings emphasize the necessity of moving beyond traditional AV evaluation metrics to ensure that AVs integrate safely and predictably into mixed-traffic environments.

The main limitation of the case study is the absence of rear-view mirrors in the VR setup (a technical limitation), requiring participants to rely on head movements. Participant feedback indicated that the lack of mirrors deviated from real-world driving conditions, and the physical strain of wearing a VR headset led to fatigue. This fatigue may have reduced head movements, particularly in later trials, potentially influencing results. Additionally, since conditions were randomized, some may have been repeated toward the end, further amplifying the impact of fatigue.

Another technical limitation was the network synchronization within the multi-player simulator, potentially introducing minor delays and slight variability in vehicle positions at the moment of driver takeover. This may have introduced noise in our experiment.

Finally, while we didn't disclose the presence of AVs in our experiment, the questionnaire referred to colour-coded vehicles, potentially influencing participants by emphasizing differences and triggering perceptions of "non-normal" behaviours. To address this, future studies could implement a comparative approach, where one group is explicitly informed about the AV’s presence while the other remains unaware. 

\section{Discussion}
Traditional evaluation of interaction-aware controllers for AVs focuses on technical performance metrics, such as collision rates, task efficiency, and trajectory optimization \cite{evestedt2016interaction, sadigh2018planning,fisac2019hierarchical,schwarting2019social, ban2007game,hu2019trajectory,liu2017path,wang2015game,ward2017probabilistic,zhou2016impact}. While these metrics are essential, they fail to capture AVs' full cognitive and behavioural impact on human drivers. Here, we proposed a framework to aid in incorporating human-centred evaluation principles into evaluating AV-HDV interactions. In a case study, we demonstrated that even when an AV performs successfully on the traditional metrics, it can still induce a higher workload and lower perceived safety in human drivers. This suggest that technical efficiency in AVs may not necessarily mean they will integrate well in mixed traffic. Relying solely on technical benchmarks could thus lead to misleading conclusions about AV performance. 

Across a range of recent studies, researchers have explored how human drivers respond to AVs in mixed traffic, revealing consistent patterns of behavioural adaptation. Several field and simulator experiments \cite{soni2022behavioral, zhao2020field, gouy2014driving, rahmati2019influence, rad2021impact, trende2019investigation} show that human drivers tend to follow AVs more closely, accept shorter gaps, or feel more comfortable when interacting with AVs. This is often driven by assumptions about AV predictability or trust in their capabilities. Other studies highlight the influence of traffic design features such as dedicated lanes \cite{rad2021impact} or AV communication range \cite{guo2023study} on traffic flow and human driver behaviour. While these findings offer valuable insights into specific behaviours and scenarios, they generally focus on isolated metrics such as gap acceptance, time headways, or merging success. A smaller subset of the literature has begun to explore broader social and psychological dimensions of interaction \cite{liu2024social, siebinga2023modelling}, emphasizing the importance of communication and human-like behaviour modelling in AV design. However, even in these cases, there is limited discussion on how to systematically measure and incorporate these human factors into AV evaluation. Other studies focus more on the impact of AVs on traffic dynamics, considering both quantitative and qualitative aspects. For example \cite{ferrarotti2024autonomous} showed that AVs can reduce time and pollution levels, and human-driven vehicles benefit from optimizing AVs dynamics.
Our framework brings together fragmented existing interaction-focused approaches and systematizes them, introducing a structured and multi-dimensional approach to assessing AV-HDV interactions spanning not only technical performance but also interaction ability, effort, effect, and perception. This enables a more comprehensive understanding of how AVs influence, and are interpreted by, human drivers in complex traffic environments.

Furthermore, beyond its practical applications, our framework contributes to fundamental research on human cognition and behaviour in automation-rich environments. The findings on driver workload and perceived safety raise important questions about how humans perceive, interpret, and respond to autonomous systems. For instance, our results indicate that human drivers tend to allocate more attention to AVs than HDVs, suggesting an increased cognitive load.
Future research could explore whether increased workload in AV interactions leads to behavioural adaptations over time or if sustained cognitive effort causes long-term stress and discomfort.

Finally, while existing AV regulations and safety standards focus on behaviour in collision avoidance situations, accident rates, and compliance with traffic laws, they do not account for the cognitive burden AVs place on human drivers. This raises an important question: \textit{Should AVs be required to communicate their intentions more clearly?} In human driving interactions, subtle cues such as eye contact, turn signals, and gestures help reduce uncertainty. If AVs fail to provide similar feedback, they may disrupt traffic flow and increase driver stress. Recent studies have explored the role of external human-machine interfaces (eHMIs) in facilitating AV-to-human communication. For instance, Tran et al. \cite{tran2024evaluating} utilized a multi-pedestrian virtual reality simulator to evaluate various eHMI designs, finding that certain interfaces significantly improved pedestrians' understanding of AV intentions, leading to increased trust and safer crossing decisions. Similarly, Guo et al. \cite{guo2022external} investigated the effects of different eHMI types on pedestrian crossing behaviour, concluding that textual displays were among the most effective in conveying clear intentions. These findings suggest that well-designed eHMIs can enhance predictability and comfort for human road users. However, as Risto et al. (2017) demonstrate, vehicle movement gestures (such as early slowing, creeping forward, or stopping short) already serve as powerful non-verbal cues in driver-pedestrian interactions. Their study shows that road users interpret these patterns as purposeful and communicative, shaping decisions and signalling intent even in complex multi-party intersections. This underscores the potential of leveraging motion-based communication, either alone or in tandem with visual eHMIs, to improve AV transparency and social acceptability on the road.

Despite these promising developments, a major obstacle persists: there is still no widely adopted, structured method for analysing AV-to-human communication, and the interactional dimensions of AV behaviour remain under-researched.
In fact, integrating social insights into the AV design process remains a challenge~\cite{schieben_designing_2019,saadatnejad_are_2022,wang_social_2022}. As highlighted by Vinkhuyzen and Cefkin \cite{vinkhuyzen2016developing}, social science research groups within engineering labs often face difficulties in translating observations of human behaviour into formal models that can be implemented in AV systems. Despite these challenges, their ethnographic research has influenced Nissan's autonomous vehicle design and company discourse, underscoring the potential benefits of incorporating social considerations into AV development. Therefore, it would be highly beneficial to systematically measure and integrate AV-to-human communication strategies during the design phase, ensuring that AVs can interact with human drivers and pedestrians in ways that are both safe and socially intuitive.
This work aims to highlight this critical research gap and to encourage broader investigation into AV-human interaction by offering a structured framework to support future studies and guide more socially aware AV development.
    
\subsection{Limitations and Future Work}
While our framework advocates a holistic assessment, it also presents a challenge: how should one design and interpret the chosen metrics within each domain? Given the complexity and subjectivity of human-AV interactions, selecting the most appropriate and meaningful metrics is not straightforward. The balance between specificity and generalization requires careful consideration, as misinterpretation of results could lead to biased or incomplete assessments. Acknowledging this challenge, researchers should approach metric selection critically and refine them based on their specific use case. Our ultimate goal was not to create a “checklist”, but to encourage a more comprehensive and interaction-aware evaluation that can evolve with ongoing advancements in AV research.

In this work, we only applied our framework to one specific case study. Future research should further develop and validate the framework by applying it to a broader range of AV controllers and driving scenarios, such as lane changes, roundabouts, and urban intersections. Evaluating how different controllers influence interaction dynamics will provide a more comprehensive understanding of AV behaviour in diverse real-world contexts.

Beyond research applications, this framework raises an important question: \textit{Should human experience and perception be incorporated into AV safety assessments?} Current methodologies primarily focus on technical performance metrics such as collision rates and efficiency, often neglecting how AVs impact human drivers' cognitive load, predictability, and trust. By integrating interaction-aware metrics, this framework provides a structured approach to broaden AV evaluation, ensuring that AVs are assessed not only for their operational safety but also for their impact on human drivers’ experience and road dynamics. For example, as shown in the case study, AVs create an “invisible” risks such as increased cognitive load and decreased perceived safety, these factors should be quantified and incorporated into certification processes. This perspective could inform policy and regulatory discussions, advocating for the inclusion of human-centred metrics in AV safety assessments. Future work should explore how these insights could shape industry guidelines and regulatory frameworks, ultimately contributing to the development of safer and more human-compatible AV systems.

\subsection{Conclusion}
The integration of AVs into mixed-traffic environments presents challenges that extend beyond traditional performance evaluations. While AVs are typically assessed based on technical metrics such as efficiency, collision rates, and manoeuvre success, our findings demonstrate that these metrics alone do not capture the full complexity of AV-HDV interactions. The case study results provide clear evidence that AVs can induce higher driver workload and lower perceived safety for other drivers even when successfully executing merging manoeuvres. 

The proposed interaction-aware evaluation framework provides a structured guide for assessing AV behaviour in real-world traffic interactions. By incorporating human-centric metrics, such as workload and perceived safety, this framework offers new insights for researchers, policymakers, and AV developers. Its adaptability allows it to be applied to different driving scenarios and AV controllers, making it a valuable tool for shaping the future of AV assessment methodologies.

As AV technology continues to evolve, ensuring that AVs are not just functionally safe but also intuitively integrated into human traffic systems will be essential for successful deployment and public acceptance.

\bibliographystyle{ACM-Reference-Format}
\bibliography{mybib}

\end{document}